# FIELD EMISSION ANALYSIS IN SRF CAVITIES FOR PIP-II USING GEANT4 *

S. A. K. Wijethunga[†], A. Sukhanov, O. Napoly, K. McGee
Fermi National Accelerator Laboratory, Batavia, Illinois, USA

*Abstract*

Field emission (FE) remains a significant hurdle for achieving optimal performance and reliability in superconducting radiofrequency (SRF) cavities used in accelerator cryomodules. A thorough understanding of the generation and propagation of FE-induced radiation is therefore essential to mitigate this problem. The absence of standardized measurement protocols further complicates the comparison of radiation data across different testing phases and facilities. This highlights the need for a precise quantitative method to diagnose and analyze FE-induced radiation. Such efforts could prove beneficial for improving cavity preparation and cleanroom assembly techniques during the prototype and production stages of Fermilab's Proton Improvement Plan-II (PIP-II) project. This study presents the initial steps of detailed Geant4 simulations aimed at analyzing FE-induced radiation in the low-beta 650 MHz 5-cell elliptical (LB650) cavity. Our goal is to combine these results with radiation diagnostics to enhance diagnostic accuracy and optimize detector positioning. This integrated approach ultimately aims to improve the preparation, assembly, and testing procedures for PIP-II SRF cavities, ensuring the delivery of FE-free cryomodules.

## INTRODUCTION

Superconducting radiofrequency (SRF) cavities are central to modern accelerators, where achieving high accelerating gradients while maintaining low losses is essential for meeting performance goals. Fermilab's Proton Improvement Plan-II (PIP-II) is one such large-scale project, aimed at upgrading the accelerator complex to deliver 1.2 MW of beam power to the Long-Baseline Neutrino Facility (LBNF). The centerpiece of PIP-II is a new superconducting 800 MeV linac that injects beam into the existing 8 GeV Booster. This linac employs 650 MHz, five-cell elliptical cavities to accelerate up to 2 mA peak current of $H^-$ ions in the energy range 185–800 MeV. The low-beta section (LB650, $\beta g = 0.61$) is designed to accelerate the beam from 185 MeV to 500 MeV using 33 dressed cavities distributed across 11 cryomodules [1].

Despite these advances, performance can be limited by field emission (FE), a quantum tunneling process triggered by surface contaminants such as dust, metallic flakes, adsorbed gases, or other impurities. Once emitted, electrons are accelerated by the RF fields, extracting stored energy from the cavity and thereby reducing the intrinsic quality factor ($Q_0$). When these electrons strike the cavity walls, they generate localized heating that increases the cryogenic load and produces Bremsstrahlung radiation in the form of X-rays. Because their trajectories are governed by the time-varying RF fields, the impact locations of electrons may be far from their emission sites. This nonlocal behavior complicates the identification of emitters and makes FE mitigation a persistent challenge.

X-ray detection outside the cryostat has become the standard diagnostic for FE during cavity RF testing. Radiation detectors are simple to operate, can be positioned flexibly, and provide quantitative information about the underlying FE processes. However, external measurements are inherently limited, as the detected radiation strongly depends on the spatial distribution and energy of electrons within the cavity—parameters that cannot be directly observed. Bridging this gap requires detailed simulations to connect measurable radiation signals with the underlying emission phenomena, providing insight that is critical for both diagnostics and mitigation strategies.

This study represents an initial effort to reconstruct FE-induced radiation in the LB650 cavity. The modeling approach is twofold. First, electron trajectories originating from potential emission sites were simulated in CST Microwave Studio [2], which tracked their motion in the RF fields and determined their impact locations and corresponding energies. These results were then used as inputs for Geant4 [3] simulations to model the resulting FE-induced radiation, with particular emphasis on the directionality and propagation of X-rays in Fermilab's Spoke Test Cryostat (STC). Together, these tools provide a systematic framework to identify probable emission sites, guide improvements in cavity cleaning and assembly procedures, and establish a basis for correlating radiation measurements with simulations in future cryomodule tests.

## SOURCE PARTICLE GENERATION

To simulate electron emission from potential sites, CST Microwave Studio was employed. First, the eigenmode solver was used to compute the cavity's electric and magnetic field distributions. These fields were then imported and scaled to the operating gradient of $E_{acc}$=16.4 MV/m (for a fixed RF phase). Using the CST particle tracking (TRK) solver, field-emitted electron trajectories were simulated to determine their motion in the RF fields, along with their impact locations and corresponding energies. The Fowler–Nordheim (F–N) equation, shown in Eq. (1), describes this emission as a quantum tunneling process through the surface potential barrier under a strong electric field.

---

$$I(E) = \frac{A_{FN}A_e(\beta_{FN}E)^2}{\phi} \exp\left(-\frac{B_{FN}\phi^{3/2}}{\beta_{FN}E}\right) \quad (1)$$

where $I = jA_e$ is current from the emitter, $A_e$ is effective emitting area, $E$ is instantaneous value of the surface electric field $\beta_{FN}$ field enhancement factor that considers the effect of emitter geometry (typical value 50-500 for SRF cavity), $\phi$ is the metal work function, 4.3 eV for Nb, $A_{FN} = 1.54 \times 10^{-6}$ $AeV/V^2$ and $B_{FN} = 6.83 \times 10^9$ $(eV^{-1.5})V/m$ [4]. It relates the emitted current density to the local electric field and the work function of the material, providing the standard framework for quantifying and analyzing field emission in SRF cavities. Since the iris region corresponds to the location of maximum surface electric field and is therefore most susceptible to generating emitters, iris locations were selected as emission points, as illustrated in Fig. 1 (e.g., iris 2). For the simulations, at least $5 \times 10^5$ emission points were uniformly distributed across the defined area source.

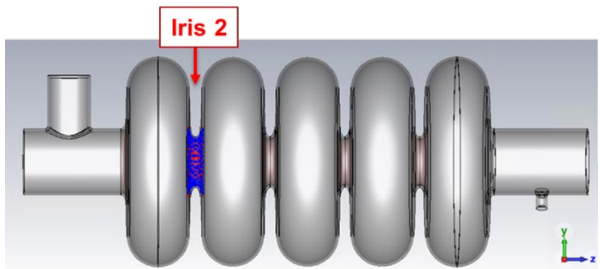

Figure 1: CST LB650 cavity model – iris 2 selected as the emission area.

While several emission models are available in CST, the field-induced model was chosen to capture the expected field dependence of electron emission. The emitted current density from the surface is then described by $J = aE^2 \exp(-b)/E$. By comparing this relation with the F-N equation, a and b were calculated as $3.58 \times 10^{-3}$ $A/V^2$ and $6.09 \times 10^8$ $V/m$, respectively, while setting $\beta_{FN} = 100$. The initial electron energy was determined by a temperature of 2 K, following the Maxwell-Boltzmann distribution. Fig. 2 shows the electron trajectories emitted from iris 2. The simulation ran for 2.6 ns, and the electron reached a maximum energy of approximately 6 MeV.

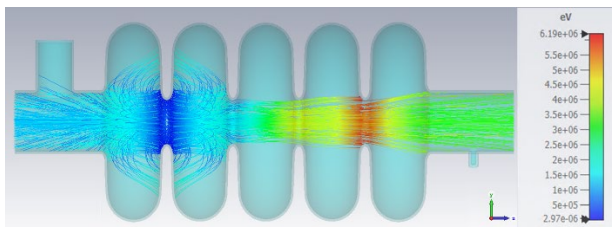

Figure 2: Electron trajectories generated from iris 2 at $E_{acc}$ = 16.4 MV/m with TRK solver.

Figure 3 shows the position and energy distributions of electrons striking the cavity surface. Approximately 3 × 10⁵ particles were collected by the cavity walls, with most impact energies below 3 MeV. These particle distributions were subsequently used as the primary input for the Geant4 simulation.

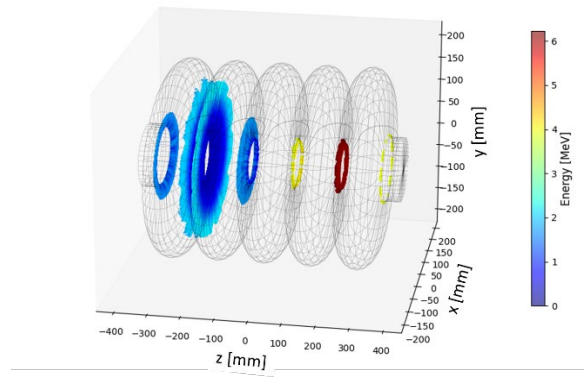

Figure 3: The position and energy distribution of electrons impacting the cavity surface.

## GEANT4 MODELLING

Geant4 is a Monte Carlo–based toolkit developed to simulate the transport and interaction of elementary particles with matter. In this work, it was used to model the radiation generated when field-emitted electrons struck the cavity walls and surrounding shielding layers. The dominant physics processes considered were bremsstrahlung, Compton scattering, and ionization, modeled using the G4EmStandardPhysics_option4 physics list, which includes multiple scattering and provides an accurate description of low-energy electromagnetic interactions [5-6].

The Fermilab Spoke Test Cryostat (STC) is used for testing SRF cavities for the PIP-II project. The cryostat consists of several shielding layers of different materials and thicknesses, as illustrated in Fig. 4 [7] and summarized in Table 1.

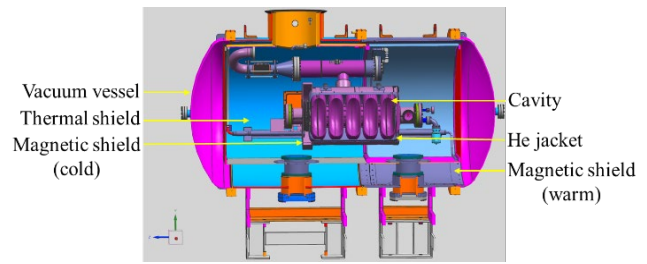

Figure 4: STC LB650 cavity assembly.

Table 1: STC Shielding Layer Details

|  | Material | Thickness [mm] |
| --- | --- | --- |
| Vacuum vessel | Stainless Steel | 12.7 |
| Magnetic shield (warm) | FeMnCrNi | 1.6 |
| Thermal shield | Al | 4.8 |
| Magnetic shield (cold) | FeMnCrNi | 1.5 |
| He jacket | Ti | 5.0 |
| Cavity | Nb | 4.0 |

For the Geant4 model, the geometry was simplified while preserving the correct material composition and layer thicknesses to capture the overall radiation pattern. The LB650 cavity was constructed using the G4Polycone class, and the surrounding shielding layers were implemented with G4Tubs using their actual dimensions and materials as shown in Fig. 5.

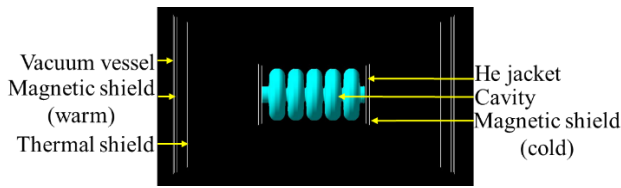

Figure 5: Geant4 STC cavity assembly.

Primary particles for the Geant4 simulations were generated from electron impact distributions obtained in the CST tracking described earlier. A total of one million electrons were produced by random resampling, and the simulations were executed on the NERSC high-performance computing facility.

Figure 6 shows the particles generated by electrons impacting the cavity surface described above, tracked as they exit the cavity volume. Secondary electrons are shown in red, while the resulting gamma-ray showers are depicted in green.

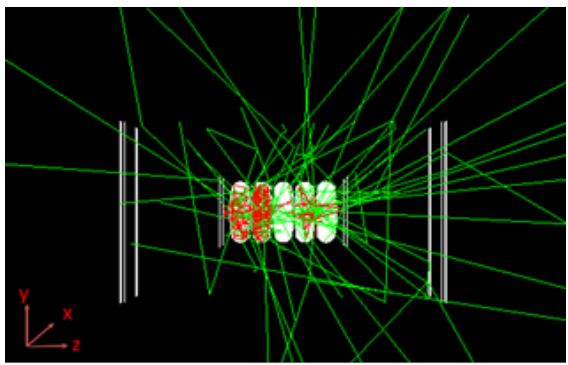

Figure 6: Particle shower generated by the electrons impacting the cavity surface shown in Fig. 3. Red indicates secondary electrons, while green represents gamma particle trajectories.

## RESULTS ANALYSIS

To analyze the simulation results, particle transport was tracked in two stages: first, as electrons exited the cavity volume, and second, as they passed through the vacuum vessel and propagated up to 2 m to the detector plane. At the end of each trajectory, particle data were recorded for post-processing.

Figure 7 illustrates the simulated gamma-ray distribution at the cavity exit: (a) a scatter plot of ϕ (in the x–y plane) versus z, and (b) and (c) histograms of photon counts along the y and x directions, respectively. The simulations indicate that 42.8% of secondary electrons and gamma rays exit the cavity volume, of which 22.8% subsequently pass through the cold magnetic shield. These results demonstrate that the shielding layers effectively absorb a large fraction of the emitted gamma radiation.

Figure 8 shows the gamma particle distribution in the STC cave up to 2 m. All the particles that exited the vacuum vessel reached the 2 m boundary.

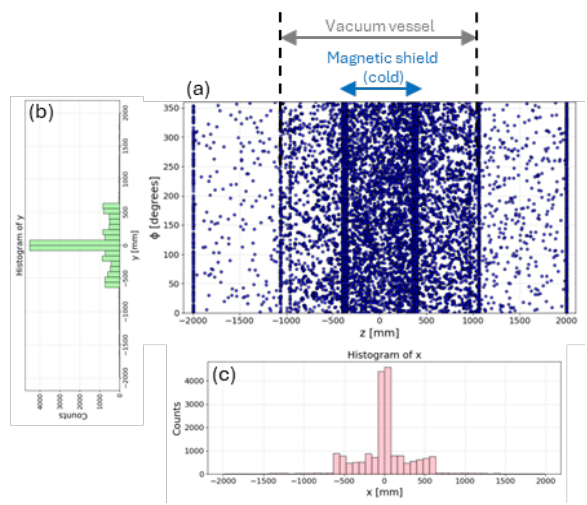

Figure 7: Simulation of gamma ray collection exiting the cavity volume, (a) scatter plot represents ϕ (in x and y direction) vs z, (b) and (c) histograms represent the photon count on the y and x directions, respectively.

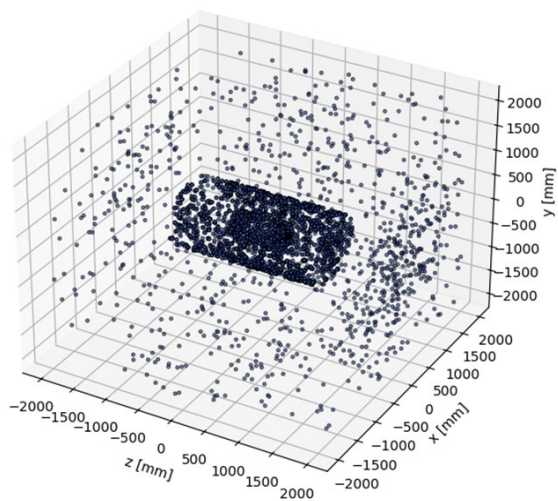

Figure 8: Illustration of the gamma particle distribution in the STC cave up to 2 m.

Figure 9 illustrates the simulated gamma-ray distribution exiting the vacuum vessel: (a) a scatter plot of y versus x–z plane, and (b) and (c) histograms of photon counts along the x and z directions, respectively. The simulations show that only about 5% of impacted particles escape the vacuum vessel as gamma particles and reach the radiation detectors.

Figure 10 presents the corresponding energy distribution of these particles, indicating that less than 1% possess energies greater than 5 MeV.

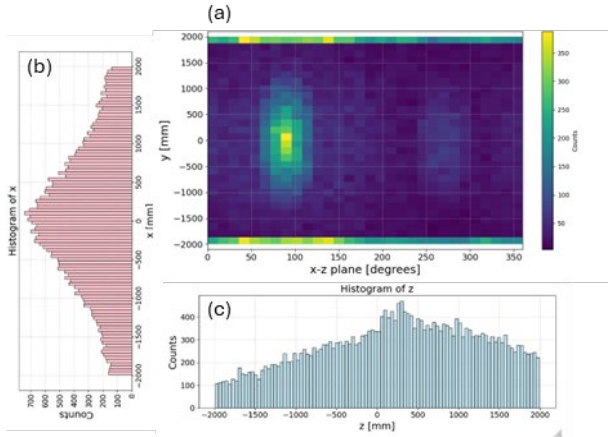

Figure 9: Simulation of gamma ray collection exiting the vacuum vessel, (a) scatter plot represents y vs x and z plane, (b) and (c) histograms represent the photon count on the x and z directions, respectively.

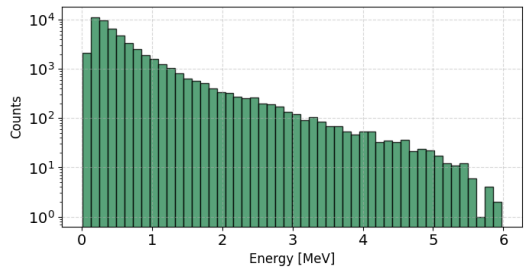

Figure 10: Emitted gamma particle energy distribution. Less than 1% of gamma particles have energy > 5 MeV.

## SUMMARY AND OUTLOOK

A simulation framework combining CST and Geant4 has been successfully developed to study field emission in LB650 cavities tested in the Fermilab STC. This work represents a first step toward understanding the directionality and propagation of FE-induced X-rays.

Future efforts will focus on systematic analyses of field emission at multiple iris locations across the full RF phase to identify the most probable emission sites. Radiation detectors will be integrated into the Geant4 model and compared with LB650 cavity measurements to enhance diagnostic accuracy and optimize detector placement. The simulation framework will be extended to SSR1, SSR2, and HB650 cavity assemblies, as well as to PIP-II cryomodules, providing a comprehensive tool for field emission characterization and mitigation in cryomodule tests.

## ACKNOWLEDGMENT

We are grateful to Soon Yung Jun (PS/PDSN) and Dinupa Nawarathne (NMSU). This research used resources of the National Energy Research Scientific Computing Center (NERSC), a Department of Energy User Facility (project m4599-2025)